\documentclass[vecphys]{svmult}
\usepackage{makeidx}         
\usepackage{graphicx}        
\usepackage{multicol}        
\usepackage[bottom]{footmisc}
\usepackage{subfigure}       

\makeindex

\begin{document}

\title*{Searches for Gravitational Waves from Binary Neutron Stars: A Review}
\author{Warren G. Anderson\inst{1} \and Jolien D. E. Creighton\inst{2}}
\institute{Department of Physics, University of Wisconsin -- Milwaukee,
P.O. Box 413, Milwaukee, Wisconsin, 53201-0413,
U.S.A. \texttt{warren@gravity.phys.uwm.edu} \and 
Department of Physics, University of Wisconsin -- Milwaukee,
P.O. Box 413, Milwaukee, Wisconsin, 53201-0413,
U.S.A. \texttt{jolien@gravity.phys.uwm.edu}}
\maketitle
\begin{abstract}%
    A new generation of observatories is looking for gravitational waves.
    These waves, emitted by highly relativistic systems, will open a new
    window for observation of the cosmos when they are detected. Among the
    most promising sources of gravitational waves for these observatories are
    compact binaries in the final minutes before coalescence. In this article,
    we review in brief interferometric searches for gravitational waves
    emitted by neutron star binaries, including the theory, instrumentation
    and methods. No detections have been made to date.  However, the best
    direct observational limits on coalescence rates have been set, and
    instrumentation and analysis methods continue to be refined toward the
    ultimate goal of defining the new field of gravitational wave astronomy.
\end{abstract}

\section{Introduction} \label{s:intro}

It is no great exaggeration to say that the advent of a new type of astronomy
is imminent. Gravitational wave astronomy is predicated on the observation of
the cosmos not with a new band of the electromagnetic spectrum, but rather via
a whole new spectrum, the spectrum of gravitational waves. As such it has the
potential to revolutionize our understanding of the Universe because it will
allow us to access phenomena which are electromagnetically dark or
obscured~\cite{Thorne1987}. The impetus for this revolution is the recent
construction and operation of a new generation of gravitational wave detectors
based on interferometry~\cite{Saulson_1994}.  Within a decade, these
interferometers will be reaching sensitivities at which gravitational wave
observations should become routine.

There are four categories of gravitational wave signals which ground-based
interferometers are currently trying to detect: quasi-periodic signals, such
as those expected from pulsars~\cite{gr-qc/0702039, gr-qc/0605028,
gr-qc/0508065, gr-qc/0410007, gr-qc/0308050}, stochastic background signals,
such as remnant gravitational waves from the Big Bang~\cite{gr-qc/0703068,
gr-qc/0703234, astro-ph/0608606, astro-ph/0507254, gr-qc/0312088}, unmodeled
burst signals, such as those that might be emitted by
supernovae~\cite{astro-ph/0703419, gr-qc/0511146, gr-qc/0507081,
gr-qc/0505029, gr-qc/0501068, gr-qc/0411027, PRD-69/102001}, and inspiral
signals, such as those from neutron star or black hole
binaries~\cite{gr-qc/0512078, gr-qc/0509129, gr-qc/0505042, gr-qc/0505041, 
gr-qc/0403088, gr-qc/0308069, gr-qc/0012010}. In this article, we will only be
concerned with the last of these searches, and in particular the search for
neutron star
binaries~\cite{gr-qc/0512078,gr-qc/0505041,gr-qc/0308069,gr-qc/0012010}.

This article is organized as follows: Since many astronomers may not be very
familiar with gravitational waves and the effort to use them for astronomy,
Sect.~\ref{s:background} is devoted to background material. This includes a
simple description of what gravitational waves are and how they are generated,
a brief history of the instruments and efforts to detect them, and some
background on the relevant aspects of neutron star binaries and the
gravitational waves they produce. Section~\ref{s:method} describes in some
detail how searches for gravitational waves from neutron star binaries have
been conducted, including a description of the data, the data analysis methods
employed, and coincidence vetoes.

Section~\ref{s:results} reviews the published upper limits that have been
placed on the rate of neutron star binary coalescence in our galactic
neighborhood by interferometric detectors.  Also included is a description of
the statistical analysis used to place these upper limits. In
Sect.~\ref{s:fp}, we discuss prospects for better upper limits and discuss
some of the astrophysics that might be done by interferometric gravitational
wave detectors when they reach a sensitivity where gravitational waves from
neutron star binaries are regularly observed.  Concluding remarks are found in
Sect.~\ref{s:conc}.

\section{Background} \label{s:background}
\subsection{Gravitational Waves} \label{ss:GWs}

One of the many remarkable predictions of Einstein's general theory of
relativity is the existence of gravitational waves (GWs)~\cite{EinsteinGR}.
Einstein himself elucidated the theoretical existence of GWs as early as
1918~\cite{EinsteinGW}. Today, however, GWs have still not been directly
measured, although the measurements of the binary pulsar PSR1913+16
\cite{TaylorWeisberg1982,TaylorWeisberg1989,WeisbergTaylor2003}, (discovered
by Hulse and Taylor and for which they won the Nobel prize) leave little doubt
that GWs do, in fact, exist. 

The fundamental factor that has led to our failure to directly measure GWs is
the exceptional weakness of the gravitational coupling constant. This causes
GWs to be too feeble to detect unless produced under extreme conditions. In
particular, gravitational waves are produced by accelerating masses (a more
exact formulation of this statement appears in Sect.~\ref{ss:BNS}). Thus, for
the highest GW amplitudes, we seek sources with the highest possible
accelerations and masses. As a result, astrophysical objects are the most
plausible sources of GWs \footnote{The exception to this statement is the
big-bang itself, which should lead to a stochastic cosmological background of
GWs, as mentioned in Sect.~\ref{s:intro}. Gravitational wave observations have
just begun to bound previously viable theoretical models of this
background~\cite{astro-ph/0608606}. We will not be considering this background
further in this review.}.  

Further restrictions on viable sources are imposed by our detectors. For
instance, for Earth based instruments, even with the best seismic isolation
technologies currently available~\cite{GiameEtAl1996,GiameEtAl2003}, noise
from seismic vibrations limits large-scale precision measurements to
frequencies above $\sim$30\,Hz. Through causality, this time-scale limitation
implies a maximum length-scale at the source of $~10^4$ km. Given the
Chandrasekhar limit on the mass of a white dwarf star~\cite{Chandrasekhar1931,
Chandrasekhar1935} and the white dwarf mass-radius relationship this is
approximately the minimum length-scale for white dwarf stars
\cite{ProvencalEtAl1998}.

If searches are restricted to objects more compact than white dwarfs, then within the bounds of current knowledge, gravitational wave
astronomy with ground-based interferometers is limited to black holes and
neutron stars as sources. According to our current understanding of
astrophysical populations, these objects are relatively rare. Thus, the
probability of finding them in our immediate stellar neighborhood are small,
and any realistic search must be sensitive to these sources out to
extragalactic distances to have a reasonable chance of seeing them in an
observation time measured in years. Is this a reasonable prospect?  The answer
is yes, but to understand why, it behooves us to first understand a bit more
about what GWs are and how they might be measured.

Gravitational waves arriving at Earth are perturbations of the geometry of
space-time.  Heuristically, they can be understood as fluctuations in the
distances between points in space. Mathematically, they are modeled as a 
metric tensor perturbation $h^{\alpha}{_{\beta}}$ on the flat spacetime
background. Linearizing the Einstein field equations of general relativity in
$h^{\alpha}{_{\beta}}$, we find that this symmetric four-by-four matrix
satisfies the the wave equation,
\begin{equation}
    \left(-\frac{{\partial}^2}{{\partial}t^2}+c^2{\nabla}^2\right)
        h^{\alpha}{_{\beta}}~=~8 {\pi}~G~T^{\alpha}{_{\beta}}, 
    \label{eq:LEFQ}
\end{equation}
in an appropriate gravitational gauge. Here ${\nabla}^2$ is the usual Laplacian
operator, $G$ is the gravitational constant, and $T^{\alpha}{_{\beta}}$ is the
stress energy tensor, another symmetric four-by-four matrix which encodes
information about the energy and matter content of the spacetime. 

From (\ref{eq:LEFQ}), it is obvious that GWs travel at $c$, the speed of
light. To understand the \emph{production} of gravitational waves by a source,
we solve (\ref{eq:LEFQ}) with the stress-energy ($T^{\alpha}{_{\beta}}$) of
that source on the right-hand-side. We will discuss this in more detail in
Sect.~\ref{ss:BNS}. First, however, we wish to consider the propagation of
gravitational waves.

For the propagation of gravitational waves, we require solutions to the
homogeneous ($T^{\alpha}{_{\beta}}=0$) version of (\ref{eq:LEFQ}). As
usual, such solutions can be expressed as linear combinations of the complex
exponential functions
\begin{equation}
    {h^{\alpha}}_{{\beta}} ={A^{\alpha}}_{{\beta}} \exp({\pm}\I k_{\mu}
        x^{\mu}).
    \label{eq:hexp}
\end{equation}
Here, ${A^{\alpha}}_{{\beta}}$ is a matrix of constant amplitudes,
$k_{\mu}=(-{\omega},\vec{k})$ is a four vector which plays the role of a
wave vector in four dimensions, and $x^{\mu}=(t,\vec{x})$ are the spacetime
coordinates. Above, and in what follows, we use the Einstein summation
convention that repeated indices, such as the ${\mu}$ in $k_{\mu}
x^{\mu}=-{\omega}t+\vec{k}{\cdot}\vec{x}$, indicate an implicit summation.  It
can be shown that for gauges in which (\ref{eq:LEFQ}) holds that 
\begin{equation}
    k_{\mu} {h^{\mu}}_{\beta}=0. 
    \label{eq:lgauge}
\end{equation}
In words, this means that the wave vector is orthogonal to the directions in
which the GW distorts spacetime, i.e. the wave is transverse. 

Since ${h^{\alpha}}_{\beta}$ is a four-by-four matrix, it has 16 components.
However, because it is symmetric, only 10 of those components are independent.
Further, (\ref{eq:lgauge}) imposes four constraints on
${h^{\alpha}}_{\beta}$, reducing the number of free components to six. One can
use remaining gauge freedom to impose four more conditions on
${h^{\alpha}}_{\beta}$. There are therefore only two independent components of
the matrix ${h^{\alpha}}_{\beta}$. Details can be found in any elementary
textbook on General Relativity, such as~\cite{Schutz1985}.

\begin{figure}[htb]
    \begin{center}
        \includegraphics[angle=-90, width=\textwidth]{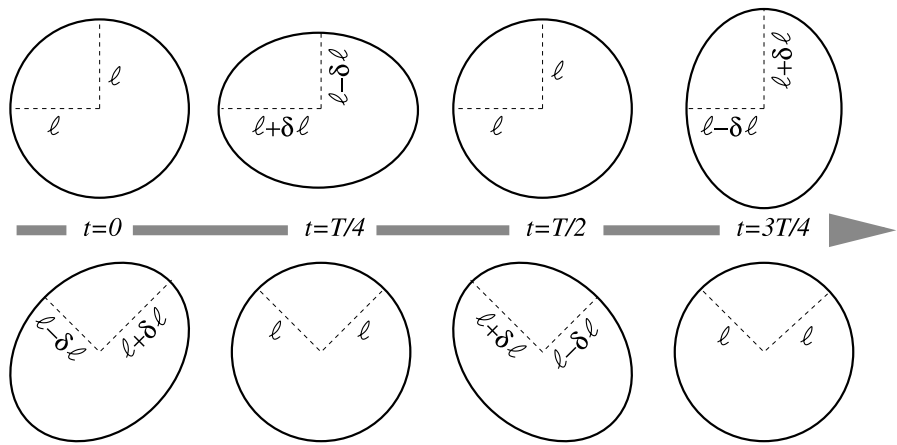}
    \end{center}
    \caption{Distortion of a ring of freely falling dust as a gravitational
    wave passes through. The wave is propagating into the page. From left to
    right are a series of four snapshots of the distortion of the ring. The
    top row are distortions due to $h_+$. The bottom row are distortions due
    to $h_{\times}$. The snapshots are taken at times $t=0$, $t=T/4$, $t=T/2$
    and $t=3T/4$ respectively, where $T$ is the period of the gravitational
    wave.  The relative phase between $h_+$ and $h_{\times}$ corresponds to a
    circularly polarized gravitational wave.}
    \label{f:rings}
\end{figure}

The two independent components of ${h^{\alpha}}_{\beta}$ are traditionally
called $h_+$ and $h_{\times}$.  These names are taken from the effect that the
components have on a ring of freely moving particles laying in the plane
perpendicular to the direction of wave propagation, as illustrated in Fig.
\ref{f:rings}. 
The change in distance between particles, ${\delta}{\ell}$, is proportional to
the original distance between them, ${\ell}$, and the amplitude of the
gravitational waves, ${A^{\alpha}}_{\beta}$. For a point source, which all
astrophysical sources will effectively be, the amplitude decreases linearly
with the distance from the source. For astrophysical source populations from
which gravitational wave emission have been estimated, the typical
gravitational wave strain, $h\;{\sim}\;2 {\delta}{\ell}/{\ell}$, at a detector
at Earth would be expected to be less than or of the order of $10^{-21}$
\cite{Thorne1987}.  Through interferometry, it is possible to measure
${\delta}{\ell}\;{\sim}\;10^{-18}$\,m. Thus, interferometers of kilometer
scales are required  to have any chance of measuring these sources. 

It might seem that the challenge of attempting to measure gravitational waves
is so daunting as to call into question whether it is worthwhile at all.
However, there are several factors which make the measurement of gravitational
waves attractive. First, astrophysical gravitational wave sources include
systems, such as black hole binaries, which are electromagnetically dark.
Gravitational waves may therefore be the best way to study such sources.
Second, since gravitation couples weakly to matter, gravitational waves
propagate essentially without loss or distortion from their source to the
detector. Thus, sources obscured by dust or other electromagnetically opaque
media may still be observed with gravitational waves. Also, interferometers
behave as amplitude sensing devices for GWs (like antennas), rather than
energy gathering devices (like telescopes), leading to a $1/r$ fall-off with
distance, rather than the more usual $1/r^2$ fall-off~\cite{Saulson_1994}.  

But perhaps the most compelling reason for pursuing the measurement of
gravitational waves is that they constitute an entirely new medium for
astronomical investigation. History has demonstrated that every time a new
band of the electromagnetic spectrum has become available to astronomers, it
has revolutionized our understanding of the cosmos. What wonders, then, await
us when we start to see the Universe through the lens of gravitational waves,
(which will surely begin to happen within the next decade as GW detectors
continue their inevitable march toward higher sensitivities)? Only time will
tell, but there is every reason to be optimistic.

In this article, we concentrate on one of the many sources of gravitational
waves for which searches are ongoing -- binary neutron star (BNS) systems. In
particular, the sensitive frequency band of ground-based interferometers, which
is approximately 40\,Hz to 400\,Hz, dictates that we should be interested in
neutron star binaries within a few minutes of
coalescence~\cite{CutlerEtAl1993}. These sources hold a privileged place in
the menagerie of gravitational waves sources that interferometers are
searching for. The Post-Newtonian expansion, a general relativistic
approximation method which describes their motion, gives us expected waveforms
to high accuracy~\cite{BlanchetEtAl1995,BlanchetEtAl1996,DamourEtAl2001}.
They are one of the few sources for which such accurate waveforms currently
exist, and they are therefore amenable to the most sensitive search algorithms
available. Furthermore, while the population of neutron star binaries is not
well understood, there are at least observations of this source with which to
put some constraints on the population~\cite{Stairs2004}. These two factor
give BNS systems one of the best (if not \emph{the} best) chance of discovery in the
near future. 

\subsection{Gravitational Wave Detectors} \label{ss:detectors}

Gravitational wave detectors have been in operation for over forty years now.
However, it is only in the last five years or so that detectors with a
non-negligible chance of detecting gravitational waves from astrophysical
sources have been in operation. The first gravitational wave detector to
operate was built by Joseph Weber~\cite{Weber1968}. It consisted of a large
cylinder of aluminum, two meters long and a meter in diameter, with
piezoelectric crystals affixed to either end. 

The fundamental idea for such detectors is that if a strong enough
gravitational wave was to pass by, that it would momentarily reduce the
interatomic distances, essentially compressing the bar, and setting it
ringing, like a tuning fork. The ringing would create electric voltages in the
piezoelectric crystals which could then be read off. Of course, as with a
tuning fork, the response of the apparatus is greatly increased if it is
driven at its resonant frequency (about 1660\,Hz, for Weber's bars).

Weber's bar was isolated from seismic and electromagnetic disturbances and
housed in a vacuum. He attributed the remaining noise in his instrument to
thermal motion of the aluminum atoms. This noise limited Weber's measurements
to strains of $h{\sim}10^{-16}$, about five orders of magnitude less sensitive
than the level now believed necessary to make the probability of detection
non-negligible. Nonetheless, by 1969, Weber had constructed two bars and had
observed coincident events in them although they were housed approximately
1000 miles apart. He calculated that his noise would create some of these
events at rates as low as one per thousand years, and subsequently published
his findings as ``good evidence'' for gravitational waves~\cite{Weber1969}.

Many groups followed up on this and subsequent claims of gravitational wave
detections made by Weber. No other group was ever able to reproduce these
observations, and it is now generally agreed that Weber's events were
spurious. Nonetheless, this launched the era of gravitational wave detectors,
as resonant mass detectors (as Weber-like bars are now called) began operating
in countries around the globe. Today, there is a network of these detectors
operated under the general coordination of the International Gravitational
Event Collaboration~\cite{AllenEtAl2000}. Technical advances have led to
considerable improvements in sensitivity over the past decades. They continue
to be rather narrow band detectors, however, and they are not the most
sensitive instruments in operation today.

\begin{figure}[htb]
    \begin{center}
        \includegraphics[angle=-90, width=0.75\textwidth]{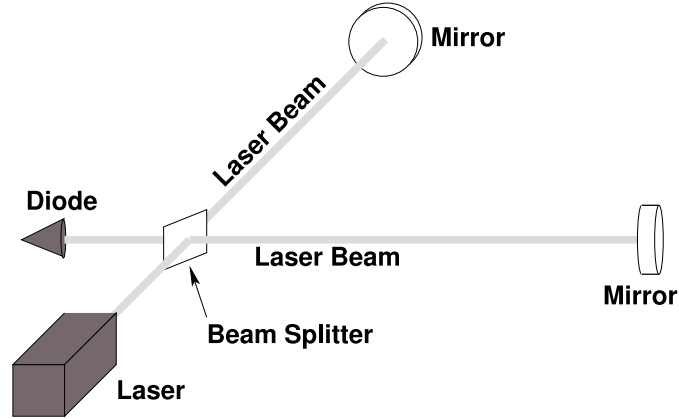}
    \end{center}
    \caption{Schematic diagram of an interferometric gravitational wave
    detector. The beam splitter is coated to allow half the light to be
    transmitted to one of the mirrors, and the other to be reflected to the
    other mirror. Real interferometric gravitational wave detectors are much
    more sophisticated, including frequency stabilization of the laser, a
    second mirror on each arm between the beam splitter and the end mirror to
    create Fabry-Perot cavities, and control feedback loops which ``lock'' the
    interferometer onto an interference fringe. Thus, rather than measuring
    the current from the photo-diode directly, the gravitational wave signals
    are encoded in the feedback loop voltages needed to maintain the lock of
    the interferometer. These and many other enhancements are necessary to
    reach the required sensitivity level.}
    \label{f:IFO}
\end{figure}

That distinction belongs to interferometric gravitational wave
detectors~\cite{Saulson_1994}. The basic components of these detectors are a
laser, a beam splitter to divide the laser light into two coherent beams,
hanging mirrors to reflect the laser beams, and a light sensing diode, as
shown
in Fig. \ref{f:IFO}.  
The mirrors act essentially as freely moving particles in the horizontal
directions. If a sufficiently strong gravitational wave with a vertical
wave-vector component impinges, it shortens the distance between the beam
splitter and one of the mirrors, and lengthens the distance between the beam
splitter and the other mirror. This is registered as a shifting interference
pattern by the light sensing diode, thus detecting the gravitational wave.

More specifically, what is measured is a quantity proportional to the
strain on the detector,
$s:=({\delta} x-{\delta} y)/{\ell}$, where ${\delta} x$ and ${\delta} y$ are
the length changes in the two equal length arms of the interferometer,
traditionally called the $x$ and $y$ arms, and ${\ell}$ is the unperturbed
length of each arm. If there were no noise, then in the special case that
$h_+$ or $h_{\times}$ was aligned with the arms, we would have ${\delta} x =
- {\delta} y = {\delta}{\ell}$ and the measured quantity would be proportional
to $2{\delta}{\ell}/{\ell}$. For a more general alignment, but still in the
absence of noise, some linear combination of $h_+$ and $h_{\times}$ is
measured 
\begin{equation}
    h(t)~:=  ~F_+({\theta},{\phi},{\psi})\,h_+(t) +
        F_{\times}({\theta},{\phi},{\psi})\,h_{\times}(t),
    \label{eq:strain}
\end{equation}
where $F_+$ and $F_{\times}$ are called \emph{beam pattern functions}. They
project the gravitational wave components, $h_+$ and $h_{\times}$, onto a
coordinate system defined by the detector, and are functions of the Euler
angles $({\theta},{\phi},{\psi})$ which relate this coordinate system to
coordinates which are aligned with the propagating GW. 

In the case where there is noise, ${\delta}x$ and ${\delta}y$ are sums of the
gravitational wave displacements and the noise displacements, so that the
detector strain is
\begin{equation}
    s(t)=h(t)+n(t)
    \label{eq:signalandnoise}
\end{equation}
where $n(t)$ is the noise contribution. If the noise component can be kept
from dominating the signal component, there is a reasonable chance that the
gravitational wave can be detected. For a more complete and detailed account
of interferometric detection of gravitational waves, as well as many other
aspects of gravitational wave physics, we recommend~\cite{Thorne1987}.

The idea of using an interferometer as a gravitational wave detectors was
explored repeatedly but independently by several different researchers over
approximately 15 years
\cite{Pirani1956,GertsenshteinPustovoit1962,WeberUnpublished,Weiss1972}. The
first prototype of an interferometric gravitational wave detector was built by
one of Weber's students, Robert Forward, and collaborators in 1971
\cite{MossEtAl1971,Forward1978}. The advantages of this idea were immediately
understood, but, as mentioned above, an understanding also emerged that
kilometer-scale interferometry would be needed. Thus, these detectors needed
to be funded, built and operated as coordinated efforts at the national or
international level.

To date, there have been six large-scale (100\,m plus) interferometers operated
at five sites. Three of the these are located in the United States and
constitute the Laser Interferometer Gravitational-wave Observatory
(LIGO)~\cite{Abramovici_et_al_1992,Barish_Weiss_1999}. There is one four
kilometer instrument at each of the LIGO-Hanford site in Washington
State~\cite{LHOweb} and LIGO-Livingston site in Louisiana~\cite{LLOweb} (H1
and L1 respectively), and a two kilometer instrument, housed in the same
enclosure, at the LIGO-Hanford Observatory (H2). The other interferometers are
the three kilometer Virgo instrument, built in Italy by a French-Italian
collaboration~\cite{AcerneseEtAl2002}, the 600 meter German-English
Observatory (GEO600) in Germany \cite{WilkeEtAl2002}, and the TAMA
observatory, a 300 meter instrument located in and funded by
Japan~\cite{TagoshiEtAl2001}. 

Other large-scale interferometers are in various stages of planning, although
funding has not been secured for them.  One of the most interesting is LISA, a
planned joint NASA-ESA mission, which would consist of two independent
million-kilometer-scale interferometers created by three satellites in solar
orbit~\cite{LISANASAweb,LISAESAweb}. The larger scale and freedom from seismic
noise will make this instrument sensitive at much lower frequencies than its
Earth-based brethren. 

Because they are currently the most sensitive gravitational wave detectors in
the world, and the ones with which we are most familiar, this article will
often use the particular example of LIGO and its methods to illustrate our
discussion. LIGO began construction in 1994 and was commissioned in 1999. It
began taking scientific data on 23 August, 2002. That data taking run, called
Science Run One (or S1 for short), ended 9 Sept., 2002. There have been four
subsequent science runs: S2 from 14 Feb. to 14 Apr. 2003,  S3 from 31 Oct.
2003 to 9 Jan.  2004, S4 from 22 Feb. to 23 Mar. 2005, and finally S5 has been
ongoing from Nov. 2005 and is expected to end in fall, 2007. At the
conclusion of S5, the LIGO interferometers are scheduled for component
enhancements which are expected to double the sensitivity of the
instrument~\cite{T060156,23TSRA}. The upgrade process is scheduled to last
approximately a year.

\begin{figure}[htb]
    \begin{center}
        \includegraphics[angle=-90,width=\textwidth]{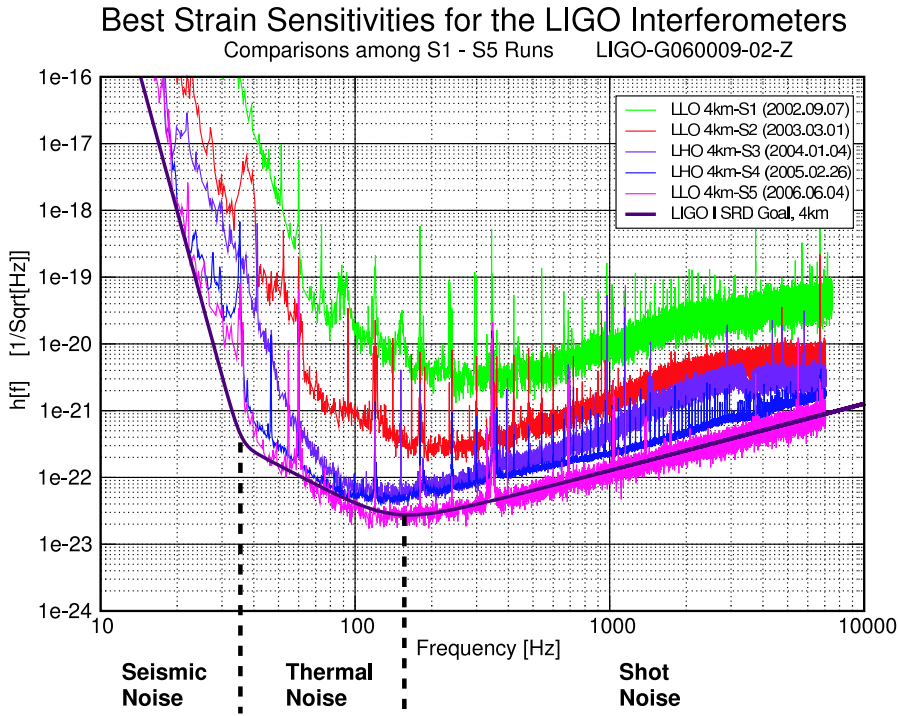}
    \end{center}
    \caption{Interferometer noise in each of the five LIGO science runs. The
    solid thick black curve is design goal of LIGO. There are two changes of
    slope indicated by dashed thick black lines. These correspond to changes
    in the dominant noise source -- the three noise regimes are marked. Over
    four years of commissioning, the noise floor was reduced by approximately
    three orders of magnitude.}
    \label{f:LIGOnoise}
\end{figure}

As with all the new interferometers, initial runs were short and periods
between them were long because scientists and engineers were still identifying
and eliminating technical and environmental noise sources which kept the
instruments from running at their design sensitivity. As designed,
interferometer sensitivity is expected to be bounded by three fundamental
noise sources~\cite{Weiss1972}. Below approximately 40  Hz, seismic noise
transmitted to the mirrors through housing and suspension dominates.  Between
approximately 40 and 200\,Hz, thermal vibrations in the suspension system for
the mirrors dominates. Finally, above approximately 200\,Hz, the dominant
contribution comes from photon shot noise associated with counting statistics
of the photons at the photodiode.  These three fundamental noise regimes,
which contribute to the ``noise floor'' of a detector, are described in the
caption of Fig.~\ref{f:LIGOnoise}.

Apparent in Fig. \ref{f:LIGOnoise} is the remarkable progress that has been
made in lowering the noise floor and the other noises superimposed upon it in
going from S1 to S5. Indeed, at S5, LIGO matches or exceeds its design
specification over most frequency bands, including the most sensitive region
from approximately 100--300\,Hz. When S5 is over and S5 data is analyzed, it
will provide by far the best opportunity to date to detect gravitational
waves.

The degree to which improvements in the noise floor such as those in
Fig.~\ref{f:LIGOnoise} correspond to increase detection probabilities depend
on the shape of the noise floor and the signal being sought. For neutron star
binary inspiral signals, one can devise a figure of merit called the inspiral
range. This is the distance, averaged over all sky positions and orientations,
to which an instrument can expect to see a 1.4-1.4\,M$_{\odot}$ binary with
signal-to-noise ratio (defined below) of 8.  For S2, typical inspiral ranges
for L1 were just below 1\,Mpc. For the LIGO design curve, the inspiral range
is approximately 14\,Mpc~\cite{Finn2001}. 

\subsection{Neutron Star Binaries as Gravitational Wave Emitters} \label{ss:BNS}

In the preceding subsections, we discussed the propagation and detection of
gravitational waves. It is perhaps useful now to say a few words about the
production of gravitational waves before diving into the particular sources of
interest for this article, binary neutron star systems.

We have discussed solutions to the homogeneous ($T^{\alpha}{_{\beta}}=0$)
version of (\ref{eq:LEFQ}), which describe the propagation of GWs in the far
wave zone.  The generation of gravitational waves requires a source, and is
therefore described by (\ref{eq:LEFQ}) with $T^{\alpha}{_{\beta}}{\neq}0$. In
fact, to lowest order, the relevant property of the source is the mass density
of the source, ${\rho}(t,\vec{x})$, or more specifically, the \emph{mass
quadrupole moments}.

The relevant quadrupole moments depend on the direction from the source at
which the gravitational wave is detected. In the limit of a negligibly
gravitating source, the moments are given by the integrals
\begin{eqnarray}
    \mathcal{J}_+(t)&:=&\frac{1}{2}\,{\int}{\rho}(t,\vec{x})
        \left[x^2-y^2\right]d^3x,\\
    \mathcal{J}_{\times}(t)&:=&{\int}{\rho}(t,\vec{x})\; x\, y \;d^3x.
    \label{eq:quadrupole}
\end{eqnarray}
Here, $\vec{x}=\{x,y,z\}$ are Cartesian coordinates centered on the source
with the $z$-axis being defined by the direction to the detector and $t$ is
time.  In terms of these integrals, the relevant components of the solution to 
(\ref{eq:LEFQ}) with source is
\begin{equation}
    h_+(t)=\frac{2}{r}\frac{G}{c^4} \frac{{\partial}^2}{{\partial} t^2}\mathcal{J}_+(t-r),
    \label{eq:quadform}
\end{equation}
and likewise for $h_{\times}(t)$.  Here $r$ is the distance between the source
and the detector.

While higher order multipole moments of the mass distribution can contribute
to the radiation, for most systems the quadrupole will dominate. Further, the
mass monopole and dipole moment will not contribute any gravitational waves.
Thus, such events as a spherically symmetric gravitational collapse and
axially symmetric rotation do not emit any gravitational radiation. On the
other hand, a rotating dumbell is an excellent emitter of gravitational waves,
making binary systems potentially amongst the brightest emitters of
gravitational waves in the Universe.

In general, the better the information about the gravitational wave signal,
the better will be our chances of detecting it. This, in turn, prods us to
find the best possible model of the dynamics of a GW emitting system. For the
gravitational wave sources suitable for Earth-based interferometric detectors,
this proves to be difficult in general. For instance, the physics of
core-collapse supernovae is not understood at the level of detail needed to
accurately predict gravitational waveforms. This is also true of colliding
black holes or neutron stars, which should emit large amounts of energy in
gravitational waves.  In fact, there are only two sources for ground-based
interferometers, inspiralling binary neutron stars and perturbed intermediate
mass (${\sim} 100$ M$_{\odot}$) black holes, for which the physics is well
enough understood that it is believed that the waveforms calculated provide a
high degree of confidence for detection\footnote{To be fair, although the
source mechanics of rotating neutron stars, like pulsars, are not well
understood, the gravitational waves they produce if there is some asymmetry
about the axis of rotation will be quasi-periodic. We therefore do not need to
model these sources well to search for their gravitational waves.  Similarly,
although a stochastic background of gravitational waves in intrinsically
unknowable in detail, a search for these waves can be optimized if their
statistical properties are known.}.

Of these two potential sources, we at least know with great certainty that
neutron star binaries, like the Hulse-Taylor binary,
exist~\cite{Stairs2004,Hulse1994,Taylor1994}. According to current thinking,
these binaries are formed from the individual collapse of binaries of main
sequence stars~\cite{KalogeraEtAl2006}. If one can establish a population
model for our galaxy, therefore, one should be able to deduce the population
outside the galaxy by comparing the rate of star formation in our galaxy to
elsewhere. One measure of star formation is blue light luminosity of galaxies
(corrected for dust extinction and
reddening)~\cite{KalogeraEtAl1991,Phinney1991}. Thus, once a galactic
population model is established, given a noise curve for an instrument it is
possible to estimate the rate at which gravitational waves from binary
neutrons star (BNS) systems out to extragalactic distances will be seen.

To develop a population model, astrophysicists use Monte Carlo codes to model
the evolution of stellar binary
systems~\cite{BelczynskiEtAl2002,BelczynskiEtAl2005}. To cope with
uncertainties in the evolutionary physics of these systems, a great many
(${\sim}$30) parameters must be introduced into the models, some of which can
cause the predicted rates to vary by as much as two orders of magnitude.  This
can be reduced somewhat by feeding what is known about BNS systems from
observation into the models~\cite{O'ShaughnessyEtAl2006}. However, only seven
BNS systems have been discovered in the galactic disk. Furthermore, the most
relevant for detection rates with interferometric GW detectors are the four
BNS which are tight enough to merge within 10\,Gyrs, since, as mentioned
above, only those within minutes of merger can be observed.  Thus, there is
not much information to feed into these models, and estimates can still vary
widely. The current best estimate is that LIGO should now be able to observe
of the order of one BNS system approximately every hundred years, although
uncertainties extend this from a few per thousand years to almost one every
ten years~\cite{KalogeraEtAl2006}. An improvement in LIGO's ability to detect
BNS systems by a factor of three would therefore raise the most optimistic
case to almost one BNS signal per year.

\begin{figure}[htbp]
     \centering
     \subfigure[Second-order restricted post-Newtonian waveforms from a
     1.4--1.4\,M$_{\odot}$ neutron star binary at 1\,Mpc]{
          \label{f:PNWaveforms}
          \includegraphics[angle=-90,width=\textwidth]{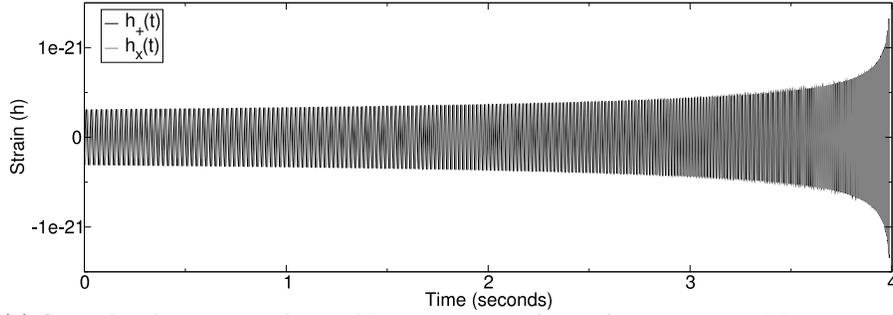}}\\
     \vspace{.2in}
     \subfigure[Closeup of above waveforms]{
           \label{f:PNWaveformsClose}
           \includegraphics[angle=-90,width=\textwidth]{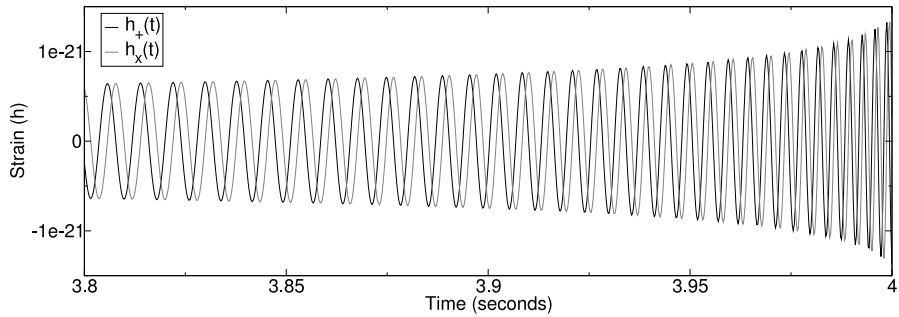}}
     \caption{Second-order restricted post-Newtonian $h_+$ and $h_{\times}$
     waveforms from an optimally located and oriented 1.4--1.4\,M$_{\odot}$
     neutron star binary system at 1\,Mpc. The top panel, (a), shows
     approximately the portion (with frequency sweeping through ${\sim}$
     40--400\,Hz) of the waveform visible to a ground-based interferometric GW
     detector. The bottom panel, (b), shows a closeup of the last 0.2\,s of
     panel (a). From the closeup, we see that the waveforms are sinusoids
     sweeping upward in both frequency and amplitude as the binary companions
     inspiral toward one another. The two phases of the gravitational wave,
     $h_+$ and $h_{\times}$, are 90$^{\circ}$ out of phase. Thus, this is a
     circularly polarized gravitational wave.}
     \label{f:BNSWaves}
\end{figure}

With projected rates this low, it is essential to search for BNS systems with
the highest possible detection efficiency. Since, in general, the more
information that can be fed into the detection algorithm, the more efficient
it will be, it is important to have a high accuracy theoretical prediction for
the waveform. In the case of BNS systems, this prediction is provided by the
restricted post-Newtonian approximation
\cite{BlanchetEtAl1995,BlanchetEtAl1996,DamourEtAl2001}. The post-Newtonian
formalism uses an expansion in orbital velocity divided by $c$, the speed of
light.  Since this is a slow motion approximation, and the orbital velocity of
the binary constantly increases during inspiral, this approximation becomes
worse as the binary evolves. However, the accuracy is still good for BNS
systems when they are in the interferometric detection band. Furthermore, we
need only deal with circular orbits since initially eccentric orbits are
circularized through the GW emission process~\cite{Peters1964}. Finally, one
can safely ignore spin terms~\cite{Apostolatos1995} and finite size
effects~\cite{BildstenCutler1992}. Figure~\ref{f:BNSWaves} shows the
post-Newtonian prediction for a neutron star binary in the sensitive frequency
band of an interferometric detector.

The general features of this waveform are easily understood. From 
(\ref{eq:quadrupole}), it is apparent that the mass quadrupole moment will be
proportional to the square of the orbital radius, $a^2$. The second time
derivative of this quadrupole moment therefore goes as $a^2{\omega}^2$, and
from (\ref{eq:quadform}) we have that $h{\sim}a^2{\omega}^2$. For a
Keplerian orbit, which relativistic orbits approximate when the Post-Newtonian
approximation is valid, ${\omega}^2{\sim}a^{-3}$ and therefore
$h{\sim}a^{-1}$. Thus, when the binary emits GWs, which carry away orbital
energy, it experiences a progressive tightening, leading to an increased
frequency and amplitude. If the two neutron stars are of equal mass, then the
quadrupole configuration repeats itself every twice every orbital period, and
the gravitational wave frequency is twice the orbital frequency. Because the
two masses are comparable in every BNS system, the dominant frequency
component is always at twice the orbital frequency.

\section{Search Method}\label{s:method}
\subsection{Interferometric Data}\label{ss:data}

As was mentioned in the caption of Fig. \ref{f:IFO}, gravitational wave
interferometers are not the simple Michelson interferometers portrayed in that
figure. Rather, they have a number of important and sophisticated refinements
to that basic configuration, all designed to increase instrument sensitivity
and reliability \cite{Saulson_1994,S1LIGOIFOs2004}. For the purpose of
understanding the data, the most important enhancement is that, when the IFOs
are operational, their arm-lengths are held fixed, or \emph{locked}. This is
done by measuring the light at the antisymmetric port (the diode in Fig.
\ref{f:IFO}) and applying feed-back controls to the mirrors through magnetic
couplings \cite{FritschelEtAl1998,FritschelEtAl2001}. By measuring the
feed-back loop voltage, it is possible to monitor how much the arm-lengths
would have changed had the mirrors been free. This allows the suppression of
pendulum modes which are either spurious or which might damage the optics if
driven at resonance.  The feedback control voltage is therefore the
gravitational wave channel of the interferometer.

In order to convert this voltage into detector strain, the frequency dependent
transfer function between these two quantities must be applied. The overall
shape of this transfer function in the frequency domain is determined by a
model of the instrument, however, the overall scaling of the transfer
function must be measured. This can be done, for instance, by driving the
mirrors at specific frequencies and measuring the response of the control loop
voltage~\cite{AdhikariEtAl2003}. At the frequencies at which these calibration
signals are injected, one can see sharp line features in the noise spectra of
the interferometers. These are some of the lines found in Fig.
\ref{f:LIGOnoise}. Other lines are caused by resonances in the wires
suspending the mirrors. Lines at multiples of 60\,Hz are caused by coupling
of the electronic subsystems to the power grid mains. Fortunately, for the BNS
search, these lines are not very problematic because the signal sweeps through
frequencies and therefore never corresponds to one of these noisy frequencies
for long.

As well as the gravitational wave channel, there are usually a myriad of other
channels monitoring the physical environment of the interferometer (seismic,
magnetic, etc) and the internal status of the instrument (mirror alignment,
laser power, etc). These can be used to veto epochs of gravitational wave data
which are unreliable.  Furthermore, if one has access to simultaneous data
from multiple interferometers (like LIGO does with its three interferometers,
all of which have similar alignment\footnote{The detectors in Washington are
somewhat misaligned with the detector in Louisiana due to the curvature of the
Earth.} and two of which are co-located), most false alarms can be eliminated
by requiring coincidence of received signals in all interferometers.

Although noise from an interferometer that is functioning well is primarily
Gaussian and stationary, there are occasional noise excursions, called
\emph{glitches}. The causes of some of these glitches are not apparent in the
auxiliary channels. These constitute the primary source of background noise in
the search for signals from neutron star binaries. 

\subsection{Matched Filtering and Chi-Squared Veto}\label{ss:filtering}

When the expected signal is known in advance and the noise is Gaussian and
stationary, the optimal linear search algorithm is \emph{matched
filtering}~\cite{WainsteinZubakov}. The idea behind matched filtering is to
take the signal, and data segments of the same length as the signal, and treat
them as members of a vector space. As with any two vectors in a vector space,
the degree to which the signal and a data vector overlap is calculated using
an inner product. 

To be more precise, consider detector strain $s(t)$ and a signal $h(t)$ that
lasts for a duration of $T$.  If the signal arrives at the detector at time
$t_0$, then the detector strain can be written
\begin{equation}
    s(t)=\left\{  
    \begin{array}{lc}
        h(t-t_0)+n(t), ~ ~ ~ ~ ~& t_0<t<t_0+T\\
        n(t), & \mbox{otherwise}
    \end{array}
    \right. 
    \label{eq:h(t)}
\end{equation}
where $n(t)$ is the detector noise. For this paragraph, we will assume that,
apart from being stationary and Gaussian, $n(t)$ is white (same average power
at all frequencies) for simplicity. Then, the matched filter output,
${\zeta}(t)$, is given by
\begin{equation}
    {\zeta}(t)= 2\,{\int}_0^Th({\tau})\,s(t+{\tau})\,d{\tau}.
    \label{eq:tdfilter}
\end{equation}
At time $t=t_0$, we have 
\begin{equation}
    {\zeta}(t_0)= 2\,{\int}_0^T h^2({\tau})\, d{\tau} 
        + 2\,{\int}_0^T h({\tau})\,n(t_0+{\tau})\, d{\tau}.
    \label{eq:tdmatch}
\end{equation}
Let us denote the first and second integrals in (\ref{eq:tdmatch}) by
$I_1$ and $I_2(t_0)$ respectively. Clearly, the integrand of $I_1$ is
deterministic and positive everywhere. However, the integrand of $I_2$ is
stochastic. The average of $I_2$ over all noise realizations vanishes. In
other words, on average ${\zeta}(t_0)=I_1$ when there is a signal starting at
time $t_0$. On the other hand, when there is no signal, ${\zeta}(t)=I_2(t)$.
Denoting the standard deviation of $I_2$ over all noise realizations by
${\sigma}$, we define the \emph{signal-to-noise ratio} (SNR) for the data to
be
\begin{equation}
    {\varrho}(t):=|{\zeta}(t)|/{\sigma}.
    \label{eq:SNRdef}
\end{equation}
Clearly, at time $t_0$ the expected value of the SNR is
${\varrho}(t_0)=I_1/{\sigma}$.  Thus, if the signal is strong enough that
$I_1$ is several times larger than ${\sigma}$, there is a high
statistical confidence that it can be detected.

In practice, it is preferable to implement the matched filter in the frequency
domain. Thus, rather than a stretch of data $s(t)$, one analyzes its Fourier
transform
\begin{equation}
    \tilde{s}(f)={\int}_{-{\infty}}^{\infty}\,e^{-2{\pi} \I f t}\;s(t)\, dt,
    \label{eq:FT}
\end{equation}
where $f$ labels frequencies. This has several advantages: first, it allows
for the non-white noise spectrum of interferometers (cf.
Fig.~\ref{f:LIGOnoise}) to be more easily handled. Second, it allows the use
of the stationary phase approximation to the restricted post-Newtonian waveform
\cite{Thorne1987,SathyaprakashDhurandhar1991}, which is much less
computationally intensive to calculate, and accurate enough for
detection~\cite{DrozEtAl1999}. Third, it allows one to easily deal with one of
the search parameters, the unknown phase at which the signal enters the
detector's band.

In the frequency domain, the matched filter is complex and takes the form
\begin{equation}
    z(t)=x(t)+iy(t)=4{\int}_0^{\infty}\,\frac{\tilde{s}^*(f)\tilde{h}(f)}{S_n(f)}\;
          e^{2{\pi} \I f t}\; df,
    \label{eq:fdmatch}
\end{equation}
where $S_n(f)$ is the one-sided noise strain power spectral density of the
detector and the $*$ superscript denotes complex conjugation. It can be shown
that the variance of the matched filter due to noise is
\begin{equation}
    {\sigma}^2=4{\int}_0^{\infty}\,\frac{\tilde{h}^*(f)\tilde{h}(f)}{S_n(f)}\;df.
    \label{eq:mfvar}
\end{equation}
In terms of $z(t)$, the SNR is given by
\begin{equation}
    {\varrho}(t)=|z(t)|/{\sigma}.
    \label{eq:SNRfreq}
\end{equation}
Note that ${\sigma}$ and $z(t)$ are both linear in their dependence on the
signal template $h$. This means that the SNR is independent of an overall
scaling of $h(t)$, which in turn means that a single template can be used to
search for signals from the same source at any distance. Also, a difference of
initial phase between the signal and the template manifests itself as a change
in the complex phase of $z(t)$. Thus, the SNR, which depends only on the
magnitude of the matched filter output, is insensitive to phase differences
between the signal and the template.

Equations (\ref{eq:fdmatch}-\ref{eq:SNRfreq}) tell us how to look for a signal
if we know which signal to look for. However, in practice, we wish to look for
signals from any neutron star binary in the last minutes before coalescence.
Because, as mentioned above, finite-size effects are irrelevant, a single
waveform covers all possible equations of state for the neutron stars.
Likewise, as stated above, the spinless waveform will find binaries of neutron
stars with any physically allowable spin. Further, as just discussed, a single
template covers all source distances and initial signal phases. However, a
single template does not cover all neutron star binaries because it does not
cover all masses of neutron stars.

Population synthesis models for neutron star binaries indicate that masses may
span a range as large as ${\sim}$1--3\,M$_{\odot}$. Since mass is a continuous
parameter, it is not possible to search at every possible mass for each of the
neutron stars in the binaries. However, if a signal is ``close enough'' to a
template, the loss of SNR will be small. Thus, by using an appropriate set of
templates, called a \emph{template bank}, one can cover all masses in the
1--3\,M$_{\odot}$ range with some predetermined maximum loss in SNR
\cite{Owen1996,OwenSathya1999}.  The smaller the maximum loss in SNR, the
larger the number of templates needed in the bank. Typically, searches will
implement a template bank with a maximum SNR loss of 3$\%$, which leads to
template banks containing of the order of a few hundred templates (the exact
number depends on the noise spectrum because both $z(t)$ and ${\sigma}$ do,
and therefore the number of templates can change from epoch to epoch).

When the noise is stationary and Gaussian, then matched filtering alone gives
the best probability of detecting a signal (given a fixed false alarm rate).
However, as mentioned earlier, gravitational wave interferometer noise
generically contains noise bursts, or glitches, which provide a substantial
noise background for the detection of binary inspirals. It is possible for
strong glitches to cause substantial portions of the template bank to
simultaneously yield high SNR values. It is therefore highly desirable to have
some other way of distinguishing the majority of glitches from true signals.

The method which has become standard for this is to use a \emph{chi-squared}
(${\chi}^2$) veto~\cite{Allen2005}. When a template exceeds the trigger
threshold in SNR, it is then divided into $p$ different frequency bands such
that each band should yield $1/p$ of the total SNR of the data if the high SNR
event were a signal matching the template. The sum of the squares of the
differences between the expected SNR and the actual SNR from each of the $p$
bands, that is the ${\chi}^2$ statistic, is then calculated. The advantage of
using the ${\chi}^2$ veto is that glitches tend to produce large (low
probability) ${\chi}^2$ values, and are therefore distinguishable from real
signals. Thus, only those template matches with low enough ${\chi}^2$ values
are considered triggers. 

If the data were a matching signal in Gaussian noise, the ${\chi}^2$ statistic
would be ${\chi}^2$ distributed with $2p-2$ degrees of freedom
\cite{Allen2005}.  However, it is much more likely that the template that
produces the highest SNR will not be an exact match for the signal. In this
case, denoting the fractional loss in SNR due to mismatch by ${\mu}$, the
statistic is distributed as a non-central chi-squared, with non-centrality
parameter ${\lambda}\,{\leq}\,2{\varrho}^2{\mu}$.  This simply means that
the ${\chi}^2$ threshold, ${\chi}^*$, depends quadratically on the measured
SNR, ${\varrho}$, as well as linearly on $\mu$. 

In practice, the number of bins, $p$, and the parameters which relate the
${\chi}^2$ threshold to the SNR , as well as the SNR threshold ${\varrho}^*$
which an event must exceed to be considered a trigger are determined
empirically from a subset of the data, the \emph{playground data}. A typical
playground data set would be ${\sim}10\%$ of the total data set, and would be
chosen to be representative of the data set as a whole. Playground data is not
used in the actual detection or upper limit analysis, since deriving search
parameters from data which will be used in a statistical analysis can result
in statistical bias.  Values for these parameters for the LIGO S1 and S2 BNS
analyses are given in
Table \ref{t:params}.
\begin{table}
    \centering
    \begin{tabular}{ccccc}
        \hline\noalign{\smallskip}
        Data Set~~~& ~${\varrho}^*$ & ~~$p$ & L1 ${\chi}^*$ & H1/H2 ${\chi}^*$
        \\
        \noalign{\smallskip}\hline\noalign{\smallskip}
        S1& 6.5 & $~ ~$8 & $~ ~5\,(p+0.03{\varrho}^2)$
            &$~ ~5\,(p+0.03\,{\varrho}^2)$\\
        S2& 6.0 & $~ ~$15 & $~ ~5\,(p+0.01 {\varrho}^2)$
            &$~ ~12.5\,(p+0.01\,{\varrho}^2)$\\
        \noalign{\smallskip}\hline
    \end{tabular}
    \caption{Search algorithm parameters for S1 and S2 BNS searches. These
    parameters were determined using playground data extracted from the S1 and
    S2 data sets respectively. Note that the ${\chi}^2$ threshold,
    ${\chi}^*$, is different for the Louisiana and Washington instruments in
    the S2 run.}
    \label{t:params}      
\end{table}

Finally, let us say a few words about clustering. As discussed earlier, when a
glitch occurs, many templates may give a high SNR. This would also be true for
a strong enough signal. It would be a misinterpretation to suppose that there
might be multiple independent and simultaneous signals -- rather, it is
preferable to treat the simultaneous events as a cluster and then try to
determine the statistical significance of that cluster as a whole. The
simplest strategy, and the one used thus far, is to take the highest SNR in
the cluster and perform the ${\chi}^2$ using the corresponding template.
Another possibility might have been to take the template with the lowest
${\chi}^2$ value as representative, or some function of ${\varrho}$ and
${\chi}^2$.  In fact, there is reason to believe that the last option may be
best~\cite{gr-qc/0505041}.

\subsection{Coincidence and Auxiliary Channel Veto} \label{ss:Coincidence}

Although the $\chi^2$ veto reduces the rate of triggers from glitches, some
glitch triggers survive. However, there are further tests that can be used to
eliminate them. Most importantly, if more than one interferometer is involved
in the search, one can require consistency between their triggers. LIGO is
especially well designed in this regard. Because LIGO's three detectors are
almost co-aligned, they should all be sensitive to the same signals (although
the 2km H2 will only be half as sensitive to them as the 4km instruments).
Thus, any signal that appears in one should appear in all. On the other hand,
there is no reason for glitches in one instrument to be correlated with
glitches in another, especially between a detector located in Washington state
and the Louisiana instrument.  Thus, one way to distinguish between triggers
generated by actual gravitational waves and those generated by glitches is to
demand coincidence between triggers in different
instruments~\cite{AmaldiEtAl1989,AstoneEtAl1999}.

The most fundamental coincidence is coincidence in time. The timing precision
of the matched filtering for properly conditioned interferometric data is
${\sim}$ 1\,ms. The larger effect is the time it could take the gravitational
wave to traverse the distance between detectors (i.e. the \emph{light travel
time} between them). The maximum time delay for this is also measured in ms
(e.g. 10\,ms between the Washington site and the Louisiana site for LIGO).
Thus, triggers at one site which are not accompanied by triggers at another
site within the light travel time plus 1\,ms are likely not gravitational
waves and can be discarded. For triggers from instruments which are
sufficiently well aligned, there are several other quantities for which one
could required coincidence. Of these, the only one which had been applied to
date as a trigger veto is the template which generated the trigger -- for the
LIGO S2 search the same template was required to have generated all coincident
triggers (or represent all coincident clusters of triggers) or they were
discarded.

The final hurdle that a trigger may have to overcome to remain viable is that
it not be associated with a known instrumental disturbance. Auxiliary channels
which monitor the instruments and their environment contain information about
many potential disturbances. Those channels most likely to correspond to
spurious disturbances which would be manifest in the gravitational wave
channel have been studied intensively. To date, studies have revealed that
channels which allows for safe and useful auxiliary channel vetoes are not
forthcoming for most instruments. However, for LIGO's second science run it
was discovered that a channel which measures length fluctuations in a certain
optical cavity of the L1 interferometer had glitches which were highly
correlated with glitches in that instrument's gravitational wave
channel~\cite{ChristensenEtAl2004}.  Thus, triggers which occurred within a
time window 4 seconds earlier to 8 seconds later than a glitch in this
auxiliary channel in Louisiana were also discarded for that analysis. 

\begin{figure}[htbp]
    \begin{center}
        \includegraphics[width=0.5\textwidth]{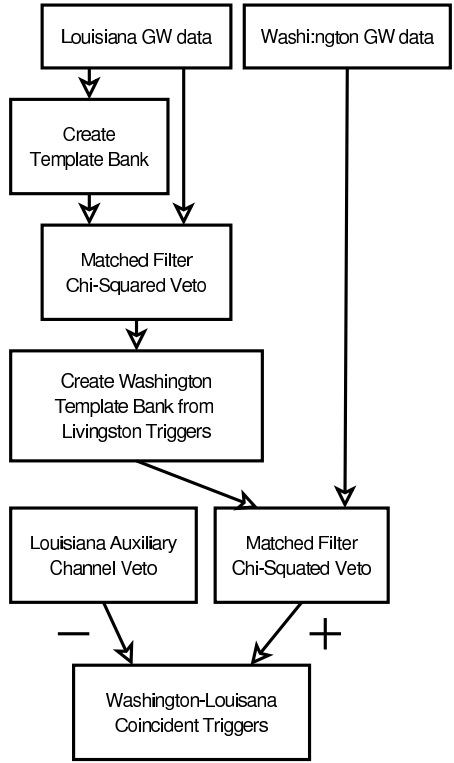}
    \end{center}
    \caption{A diagram showing a typical work-flow to look for signals from
    neutron star binaries. This is a two instrument work-flow for LIGO
    involving one detector from Washington and the Louisiana detector. It is
    similar to the one used for the S2 analysis. There would a slightly more
    complicated work-flow diagram when all three instruments were used. Note
    that the Washington data is only analyzed at times and for binary
    templates that correspond to a trigger in Louisiana, thus minimizing both
    false alarms and computing time. The + and -- signs by the two bottom-most
    arrows indicated that the addition and subtraction of coincident triggers
    respectively -- i.e. passing the SNR and ${\chi}^2$ thresholds adds
    coincident triggers, while occurring during an auxiliary channel glitch
    removes them.}
    \label{f:Pipeline}
\end{figure}

Finally, triggers which survive all of these cuts must be examined
individually to determine if they are genuine candidates for gravitational
wave signals. The flow chart for the procedure we have just describe is shown
in Fig.  \ref{f:Pipeline}. 
A more comprehensive and detailed discussion of such an algorithm can be found
in~\cite{AllenEtAl2005}.  Details of the pipelines used by LIGO for the S1 and
S2 analyses can be found in \cite{BrownEtAl2004} and \cite{Brown2005}
respectively. In the next section, we discuss how the surviving candidates are
analyzed and how upper limits are set from them.

\section{Statistics and Results}\label{s:results} 

\subsection{Background} \label{ss:background}

Once the method outlined above is completed, one is left with zero or more
coincident triggers that need to be interpreted. The pipeline, and in
particular the search parameters ${\varrho}^*$ and ${\chi}^*$, are chosen so
that the probability of missing a real signal are minimized. This means, in
practice, that the probability of having coincident triggers from glitches is
not minimized. 

These coincident triggers from glitches form a \emph{background} against which
one does a statistical analysis to determine the likelihood that there are
signals (foreground events). In order to accomplish this task, it is useful to
understand the background rate.  Ideally, one would have a model for the
background, but there is no such model for GW detector glitches. However,
there is a fundamental difference between background and foreground triggers
that can be exploited -- coincident background triggers are caused by random
glitches which happened to coincide between detectors (assuming that the
glitches in detectors are not correlated), where as foreground triggers are
coincident because they are caused by the same gravitational wave signal.
Thus, if one introduces a large enough artificial time delay between detector
data streams (e.g. greater than 11\,ms between the triggers from Louisiana and
those from Washington), the background rate should remain unchanged while the
foreground rate must vanish. 

A large number of time shifts (e.g. 40 for S2) are used to provide a great
deal of background data and thus increase confidence in the inferred
background rate.  Both the mean and variance of the background rate are
estimated. The foreground rate is then compared. If the foreground rate is
significantly higher than the mean background rate, it is possible that some
of the coincident triggers are from gravitational waves. A careful look at
each coincident trigger then ensues.  Criteria such as consistency of SNR
values, problems with auxiliary channels, and behaviour of the gravitational
wave channel are invoked to try to eliminate surviving background events. In
particular, it is known that poor instrument behavior often leads to many
consecutive glitches, and coincident triggers are much more likely during
these glitchy times.

\subsection{Upper Limits and the Loudest Event}

If no gravitational wave candidate is identified amongst the coincident
triggers, then the best one can do is set upper limits on the rate at which
such signals occur. Clearly, it is meaningless to quote a rate unless one
specifies the minimum SNR one is considering. Since triggers can have
different SNR's in each instrument, for coincident searches it is convenient
to consider a measure of the combined SNR. 

If the noise in all instruments were Gaussian, then it would be appropriate to
combine SNR's in quadrature, e.g. ${\varrho}_C^2={\varrho}_H^2+{\varrho}_L^2$,
where ${\varrho}_H$ and ${\varrho}_L$ are the SNR's at Washington and
Louisiana respectively. This would still be appropriate for non-Gaussian noise
if the rate and SNRs of glitches were the same for instruments at both sites.
However, this is not guaranteed to be the case, and if it is not, the noisier
instrument can dominate the the upper limit statistics. To avoid this problem,
the combined SNR needs to be modified slightly to 
\begin{equation}
    {\varrho}_C^2={\varrho}_H^2+{\alpha}{\varrho}_L^2,
    \label{eq:SNR_com}
\end{equation}
where ${\alpha}$ needs to be determined empirically. 

Now, let us denote the mean rate at which signals arrive at a set of detectors
with SNR ${\varrho}_C > {\varrho}_C^*$ by $\mathcal{R}$. If we model the
arrival of such signals as a Poisson process, then the probability of
detecting such a signal within time $T$ is given by
\begin{equation}
    P({\varrho}_C>{\varrho}_C^*\;;\;\mathcal{R})=
        1-e^{\mathcal{R}T{\varepsilon}({\varrho}_C^*)},
    \label{eq:Poisson}
\end{equation}
where ${\varepsilon}({\varrho}_C^*)$ is the detection efficiency, i.e. the
ratio of detected signals to incident signals at the threshold
${\varrho}_C^*$. This is not the probability of observing a trigger with
${\varrho}_C>{\varrho}_C^*$ however, because it does not account for
background triggers. If the probability of having no background event with
${\varrho}_C>{\varrho}_C^*$ is $P_b$, then the probability of observing at
least one trigger with ${\varrho}_C>{\varrho}_C^*$ is 
\begin{equation}
    P({\varrho}_C>{\varrho}_C^*\;;\;\mathcal{R},b)=
        1-P_b\,e^{\mathcal{R}T{\varepsilon}({\varrho}_C^*)}.
    \label{eq:Poisson+BG}
\end{equation}

The loudest event (i.e. the event with the maximum combined SNR,
${\varrho}_\mathrm{max}$) sets the scale for the upper limit on the rate. More
precisely, to find the 90\% frequentist upper limit, one needs to determine
the value of $\mathcal{R}$ such that there is a 90\% chance that no combined
SNR would exceed ${\varrho}_\mathrm{max}$ over the course of the run.  In other
words, one needs to solve
$0.9=P({\varrho}_C>{\varrho}_\mathrm{max}\;;\;\mathcal{R},b)$ for $\mathcal{R}$.
Doing so, we find
\begin{equation}
    \mathcal{R}_{90\%}=\frac{2.303+\ln P_b}{T{\varepsilon}({\varrho}_\mathrm{max})}.
    \label{eq:R90wBG}
\end{equation}
There are two quantities in (\ref{eq:R90wBG}) which need to be ascertained.
The first is the efficiency of the detection algorithm to signals with
combined SNR ${\varrho}_\mathrm{max}$. This is evaluated through Monte Carlo
simulations where simulated signals from the target population of neutron star
binaries (using the population models discussed in Sect.~\ref{ss:BNS}) are
injected into the data. 

The second is the background probability, $P_b$.  The most straightforward
approach might be to estimate $P_b$ using the background events resulting from
the timeshifts as described in Sect.~\ref{ss:background}. However, these
background rates are known to be subject to significant variation depending on
the details of search -- e.g. reasonable changes to the event clustering
criteria lead to different background rate estimates\cite{gr-qc/0505041}.
Since for detections one would follow up with detailed investigation of
coincident triggers anyway, the background estimates are used in a more
qualitative manner, and this variation is not an issue for detection. For an
upper limit, however, one needs a quantitative result for the background
rates, and if there is uncertainty in the value obtained, it also must be
quantified.  Failing to do so could lead to undercoverage, i.e.  a 90\% upper
limit which is below the actual rate that can be inferred from the data.
Undercoverage is considered a ``cardinal sin'' in frequentist analyses.

Therefore, rather than try to get a quantitative estimate, the standard
practice is to simply use $P_b=1$ in (\ref{eq:R90wBG}). Note that this
maximizes $\mathcal{R}_{90\%}$ with respect to $P_b$. It therefore gives a
conservative upper limit -- the actual 90\% confidence rate is certainly
lower. While undesirable, this is considered a ``venial sin'' in frequentist
statistics, and therefore far preferable to undercoverage. Furthermore, since
$P_b$ is the probability that no background coincidences will occur with SNR
above ${\varrho}_\mathrm{max}$, and since ${\varrho}_\mathrm{max}$ is the SNR
of the loudest background coincidence that actually did occur, it is
statistically unlikely that the actual $\mathcal{R}_{90\%}$ would be very much
below the one obtained by this method. This statistic is known as the
\emph{loudest event statistic}, and is discussed at some length
in~\cite{BradyEtAl2004}.

Finally, BNS rate limits are typically quoted in units of ``per Milky Way
equivalent galaxy (MWEG) per year''. This results from expressing the
efficiency in terms of effective number of Milky Way equivalent galaxies to
which the search was sensitive, $N_G$. The conversion between
${\varepsilon}({\varrho}_\mathrm{max})$ and $N_G$ is
\begin{equation}
    N_G := {\varepsilon}({\varrho}_\mathrm{max}) 
        \left( \frac{L_\mathrm{pop}}{L_G} \right),
    \label{eq:NGdef}
\end{equation}
where $L_\mathrm{pop}$ is the effective blue-light luminosity of the target
population and $L_G=9{\times}10^9\,L_{\odot}$ is the blue-light luminosity of
the Milky Way galaxy. In terms of $N_G$, the 90\% frequentist rate upper limit
is written
\begin{equation}
    \mathcal{R}_{90\%}=2.303\,{\times}\,\left( \frac{1\,\mathrm{yr}}{T}
        \right)\,{\times}\,\left( \frac{1}{N_G} \right) \mathrm{yr}^{-1}
        \mathrm{MWEG}^{-1},
    \label{eq:R90}
\end{equation}
where $T$ has units of years and $P_b=1$ has been used. This, then, is the
most common form of the upper limit that is quoted in the literature, and in
this article.

\subsection{Results} \label{ss:Results} To date, there have been five published
searches for BNS coalescence by large scale ($>$ 100\,m) interferometric
detectors\footnote{One other was done using TAMA300 and LISM, a 20\,m
prototype interferometer\cite{gr-qc/0403088} -- we will not consider this one
here since we feel that the results are essentially subsumed by the TAMA
2000-2004 result}. The first was performed by TAMA with a single
instrument~\cite{gr-qc/0012010} on the 1999 DT2 (Data Taking 2) data.
Subsequent multi-detector searches were performed by LIGO based on
S1~\cite{gr-qc/0308069} and S2~\cite{gr-qc/0505041} data and jointly by LIGO
and TAMA~\cite{gr-qc/0512078}. Most recently, TAMA has analyzed cumulatively
data taken over many Data Taking runs between 2000 and 2004. No detections
have been claimed for any analysis thus far. The results therefore have all
been observational upper limits on coalescence rates. They are summarized in
Table \ref{t:results}.
\begin{table}
    \centering
    \begin{tabular}{cccc}
        \hline\noalign{\smallskip}
        Data Set ~~~~&~~~~~ $T$ ~~~~~&~~~~~ $N_G$ ~~~~~&~~~~~
            $\mathcal{R}_{90\%}$\\
            &$_{(\mathrm{hrs})}$&&~~~$_{(\mathrm{MWEG}^{-1} \mathrm{yr}^{-1})}$\\ 
        \noalign{\smallskip}\hline\noalign{\smallskip}
        \vspace{0.2cm}
        TAMA DT2&6&-&5170\\
        \vspace{0.2cm}
        LIGO S1&236&$0.60^{+0.12}_{-0.10}$&170\\
        \vspace{0.2cm}
        LIGO S2&339&$1.34^{+0.06}_{-0.07}$&47\\
        \vspace{0.2cm}
        LIGO S2/TAMA DT8&584&$0.76^{+0.05}_{-0.06}$&49\\
        \vspace{0.2cm}
        TAMA 2000-2004&2075&-&20\\
        \noalign{\smallskip}\hline
    \end{tabular}
    \caption{Results from five searches for neutron star binaries involving
    data from large scale ($>$ 100\,m) interferometric gravitational wave
    detectors. $T$ is the observation time used for setting the upper limit,
    i.e.  the total observation time minus playground data set length minus
    time periods lost due to vetoes. In the first and last entry, TAMA did not
    convert their efficiency to $N_G$. The results quoted in these rows,
    therefore, are not in units of MWEG$^{-1}$ yr$^{-1}$, but rather in ``per
    population observed yr$^{-1}$''.  Note that where $N_G$ is used, the
    quoted rates are those from the lower bound for $N_G$, thus giving the
    most conservative upper limit.}
    \label{t:results}
\end{table}

All of these rates are substantially above the theoretically predicted rates
for these searches\cite{Kalogera2004}. Nonetheless, these are the best direct
observational limits on neutron star binaries to date. Furthermore, even at
these sensitivities, given how little is known about these gravitational wave
sources, there is a real chance, however marginal, of a serendipitous
discovery. Finally, these studies were used as a testing ground to develop the
analysis tools and methodologies that will maximize the possibility of
discoveries in future analyses.

\section{Future Prospects}\label{s:fp}
\subsection{Interferometers Now and Future} \label{ss:future}
As mentioned in the introduction, to date (Spring 2007), LIGO has completed
four science runs, S1 from 23 Aug--9 Sep 2002, S2 from 14 Feb--14 Apr 2003, S3
from 31 Oct 2003--9 Jan 2004, and S4 22 Feb-- 23 Mar 2005.  These short
science runs were interspersed with periods of commissioning and have yielded
dramatic improvements in sensitivity (see Fig.~\ref{f:LIGOnoise}).  LIGO is
now operating with its design sensitivity.  The fifth science run, S5, began
in Nov 2005, and has the goal of collecting a full-year of coincident data at
the LIGO design sensitivity.  This goal is expected to be achieved by late
summer 2007\cite{23TSRA}.  The scientific reach of S5 -- in terms of the
product of the volume of the Universe surveyed times the duration of the data
sample -- will be more than two orders of magnitude greater than the previous
searches.

A period of commissioning following the S5 run will hopefully improve the
sensitivity of LIGO by a factor of $\sim2$~\cite{23TSRA,T060156}. An
anticipated S6 run with the goal of collecting a year of data with this
``enhanced'' LIGO interferometer is projected to commence in 2009.  If S6
sensitivity is indeed doubled, enhanced LIGO will survey a volume eight times
as great as the current LIGO sensitivity.

Following S6, in 2011, the LIGO interferometers are slated for decommissioning
in order to install advanced interferometers\cite{23TSRA,AdLIGO}. These
advanced LIGO interferometers are expected to operate with 10 times the
current sensitivity (a factor of 1000 increase in the volume of the Universe
surveyed) by $\sim2014$.

In addition to LIGO, the Virgo and GEO600 observatories are also participating
in the S5 science run and are expected to undergo upgrades that will improve
their sensitivity along with LIGO. In Japan, TAMA will hopefully be replaced
by the Large Scale Cryogenic Gravitational-wave Telescope (LCGT), a
subterranean cryogenic observatory with similar performance to advanced
LIGO\cite{23TSRA}.  Having several detectors operating at the time of a
gravitational wave event will allow better determination of an observed
system's parameters.

\subsection{Future Reach and Expected Rates}\label{ss:reach}
It is possible, even before doing analyses, to get estimates of how
instruments will perform in future analyses given their noise curves. For
instance, since S2 a sequence of improvements to LIGO's sensitivity have been
made (see Fig.~\ref{f:LIGOnoise}) that have resulted in an order of magnitude
increase in the range to which it is sensitive to neutron star binary
inspiral.  Galaxy catalogs can be used to enumerate the nearby galaxies in
which BNS inspirals could be detected.  The relative contribution to the
overall rate of BNS inspirals for each galaxy is determined by its relative
blue light luminosity compared to that of the Milky Way.  Monte Carlo methods
are used to determine the fraction of BNS signals that would be detectable
from each galaxy during a particular science run. As described above, this
procedure yields a figure for sensitivity during any science run: the number
of Milky Way Equivalent Galaxies (MWEG) that are visible.  

In the case of the second LIGO science run, S2, the search was sensitive to
1.34\,MWEG.  The S2 run produced a total of 339 hours (around 0.04 years) of
analyzed data.  The inverse of the product of the number of galaxies to which
a search was sensitive times the livetime is a measure of the scientific reach
of the search.  For S2 this was
${\sim}2{\times}10^7\,\mbox{Myr}^{-1}\,\mbox{MWEG}^{-1}$.  For initial LIGO
sensitivity (which has been achieved during the current S5 science run)
Nutzman et al.~\cite{Nutzman2004} estimate the effective number of MWEG
surveyed to be ${\sim}600$ MWEG, though this number depends on the actual
sensitivity achieved during the search. Assuming that S5 produces one
year of analyzed data, the scientific reach of this search is
${\sim}2000\,\mbox{Myr}^{-1}\,\mbox{MWEG}^{-1}$.  These numbers should be
compared to estimated BNS merger rates in the Milky Way. For example, the
``reference model'' (model 6) of Kalogera et al.~\cite{Kalogera2004} quotes a
Galactic rate of between ${\sim}20$ and ${\sim}300 \,\mbox{Myr}^{-1} \,
\mbox{MWEG}^{-1}$ with a most likely rate of ${\sim}80 \,\mbox{Myr}^{-1} \,
\mbox{MWEG}^{-1}$.

Extrapolating to the future, the enhancements that are expected to double
the range of LIGO before the sixth science run could increase the
scientific reach of LIGO to ${\sim}200\,\mbox{Myr}^{-1}\,\mbox{MWEG}^{-1}$,
which would now begin to probe the expected range of BNS coalescence rates.
Since the Advanced LIGO is expected to improve the sensitivity by a factor of
10 compared to the current sensitivity, given a few years of operation a
scientific reach of ${\sim}1\,\mbox{Myr}^{-1}\,\mbox{MWEG}^{-1}$ should be
achieved. It is therefore likely that detection of BNS inspirals will become
routine during the operation of Advanced LIGO.

\subsection{BNS Astrophysics with GWs}
As discussed above, we expect that future GW observations will begin to probe
the interesting range of BNS inspiral rates over the next several years; when
advanced detectors like Advanced LIGO begin running at the anticipated
sensitivity, we expect BNS inspirals to be routinely detected, which will give
a direct measurement of the true BNS merger rate as well as observed
properties (such as the distribution of masses of the companions) of the
population.  Such constraints on the population of BNS can then be compared to
the predictions from population synthesis models, which then can give insight
into various aspects of the evolution of binary
stars~\cite{O'ShaughnessyEtAl2006}.

Design differences between current and future interferometers might also open
new avenues of investigation. One intriguing possibility involves making
measurements of neutron star equations of state using GW observations of BNS
mergers.  It is possible that advanced interferometers will be able to be
tuned to optimize the sensitivity for a particular frequency band (while
sacrificing sensitivity outside that band).  This is already being discussed
as a feature for Advanced LIGO.  This raises the possibility of tuning one of
the LIGO interferometers to be most sensitive to gravitational waves at high
frequencies, around 1~kHz, where effects due to the size of the neutron stars
is expected to be imprinted in the gravitational waveform.  Advanced
interferometers could then make a direct measurement of the ratio of neutron
star mass to radius and thereby constrain the possible neutron star equations
of state~\cite{FaberEtAl2002}.

Finally, recent evidence suggests that short hard gamma-ray bursts (GRBs)
could be associated with BNS mergers or neutron-star/black-hole binaries. Once
GW interferometers are sufficiently sensitive, a search for inspiral waveforms
in conjunction with observed GRBs could confirm or refute the role of binaries
as GRB progenitors if a nearby short GRB were identified.  If the binary
progenitor model is accepted then gravitational wave observations of the
associated inspiral could give independent estimates or limits on the distance
to the GRB.  While most GRBs will be too distant for us to hope to detect an
associated inspiral, a few may occur within the observed range.

\section{Concluding Remarks}\label{s:conc}
We have reviewed here the recently published results of searches for
gravitational waves from neutron star binaries. No detections have been made
in analyses published thus far, but this is hardly surprising given the gap
between current sensitivities and those required to reach astrophysically
predicted rates. Nonetheless, steady progress is being made in refining
instrumentation and analyses in preparation for that time in the
not-too-distant future when this gap has been closed.

It should be obvious to the reader that this has hardly been a comprehensive
description of any aspect of this search. Indeed, to provide such a
description would require at least an entire volume in itself. Notably, our
description of the mathematical theory of gravitational waves, the theoretical
and computational underpinnings of population synthesis modeling and our
discussion of interferometric design were all sketchy at best. Nor have we
delved at all into the important topics of error analysis and pipeline
validation. 

Nonetheless, we have attempted to provide a birds-eye view of the relevant
aspects of searches for gravitational waves from coalescing neutron star
binaries with enough references to the literature to provide a starting point
for readers interested in any particular aspect. And there is good reason to
believe that, in time,  gravitational wave searches will become increasingly
interesting and relevant to a ever broadening group of astronomers and
astrophysicists outside the gravitational wave community.  Already, viable (if
somewhat marginal) theoretical models have been
constrained~\cite{astro-ph/0608606}, and greater volumes of more sensitive
data are in hand. It seems that it is only a matter of time before GW
detectors begins seeing the Universe through gravitational waves. We look
forward to the opportunity to review the findings when they do.

\paragraph{Acknowledgements}\mbox{}\\
We would like to thank Patrick Brady, Duncan Brown, Matt Evans, Gabriela
Gonz\'{a}les,  Patrick Sutton and Alan Weinstein for their timely readings and
comments on this manuscript. Our understanding and appreciation of
gravitational wave physics has been enriched by our participation in the LIGO
Scientific Collaboration, for which we are grateful. The writing of this
article  was supported by NSF grant PHY-0200852 from the National Science
Foundation. This document has been assigned LIGO Document number P070053-02.

\end{document}